\documentclass[1p,sort&compress,plainnat]{elsarticlelocal}

\usepackage{graphicx,bm,bbm,enumerate,amssymb,amsmath}

\begin{document}

\date{May 14, 2010}

\author{Samuel Bieri}
\ead{samuel.bieri@a3.epfl.ch}
\author{J\"urg Fr\"ohlich}
\ead{juerg@itp.phys.ethz.ch}

\address{ETH Z\"urich, Institut f\"ur Theoretische Physik, Wolfgang-Pauli-Strasse 27, 8093 Z\"urich, Switzerland}

\title{Physical principles underlying the quantum Hall effect}

\begin{abstract}
In this contribution, we present an introduction to the physical principles underlying the quantum Hall effect. The field theoretic approach to the integral and fractional effect is sketched, with some emphasis on the mechanism of electromagnetic gauge anomaly cancellation by chiral degrees of freedom living on the edge of the sample. Applications of this formalism to the design and theoretical interpretation of interference experiments are outlined.
\end{abstract}


\maketitle
\date{May 14, 2010}

\section{Introduction}

The work reviewed in this contribution has been carried out in various collaborations, during the years 1989 - 2000 and 2008/2009 \cite{FrohlichKing89, FrohlichGabbiani90, FrohlichKerler91, FrohlichStuder93,FrohlichZee91,FrohlichThiran94, FrohlichStuderThiran95, FrohlichStuderKerlerThiran95, FrohlichPedrini2000, FrohlichPedrini2001, FrohlichPedriniSchweigertWalcher01, LevkivskyiBoyarskyFrohlichSukhorukov09, BoyarskyCheianovFrohlich09}. A useful classical reference on the quantum Hall effect is \cite{GirvinPrange87}.

\begin{figure}[!h]
\center
\includegraphics[width=0.6\textwidth]{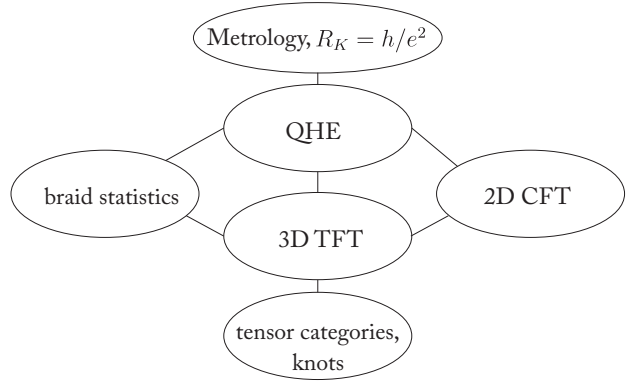}
\caption{The quantum Hall effect is related to metrology, as well as to various fundamental physical and abstract mathematical concepts.
\label{fig:overview}
}
\end{figure}

The reason the quantum Hall effect (QHE) is relevant to the subject of this colloquium, {\it metrology}, lies in the circumstance that it yields a highly precise experimental value for the {\it von Klitzing constant}
\begin{equation}
  R_K = \frac{h}{e^2}\, .
\end{equation}
This constant plays a fundamental role in the QHE: The Hall conductance of a two-dimensional incompressible electron gas (2DEG) exhibiting the QHE turns out to be an integral or rational multiple of $R_K^{-1}$. Its significance for metrology is clearly an important aspect of the QHE. Apart from that, the QHE is a fascinating phenomenon, because its theoretical description is related to quite fundamental and abstract concepts in mathematics and theoretical physics, such as fractional- or braid statistics, tensor categories, knot theory, 2D conformal field theory (CFT), and 3D topological field theory (TFT); see Fig.~\ref{fig:overview}. In these notes, we present a short introduction to some of the concepts underlying the theory of the QHE. We also provide a list of important references, with emphasis on our own contributions.

\subsection{Remarks on history}
An overview of the history of the quantum Hall effect can be found, e.g., in Ref.~\cite{jeanneret2004}. Here, we just list some important scientific milestones.

\noindent {\bf 1879} {\it Edwin Hall} discovers what is now called the {\it classical} Hall effect. Later, this discovery reveals that the electric current in some semi-conductors is carried by {\it holes}.

\noindent {\bf 1966} {\it Fowler et al.} investigate, for the first time, a two-dimensional electron gas~(2DEG) at low temperature in a strong magnetic field in a Silicon heterostructure (MOSFET).

\noindent {\bf 1975} {\it Kawaji et al.} observe a dissipationless state in a Si-MOSFET device.

\noindent {\bf 1978} Hall plateaux are observed by {\it Englert} and {\it von Klitzing}.

\noindent {\bf 1980} {\it von Klitzing} realizes that the heights of the plateaux in the Hall conductance are quantized in integral multiples of the constant $R_K^{-1}$ \cite{Klitzing80}.

\noindent {\bf 1982} {\it Tsui}, {\it St\"ormer}, and {\it Gossard} discover the {\it fractional quantum Hall effect} in GaAs-AlGaAs heterostructures \cite{TsuiStormerGossard82}.

\noindent $\geq$ {\bf 1982} {\it Laughlin} and followers \cite{Laughlin81,Laughlin83,Halperin82,Haldane83,Morf86} propose theoretical explanations of the fractional QHE.

\section{What is the quantum Hall effect?}

\begin{figure}[!h]
\center
\includegraphics[width=0.6\textwidth]{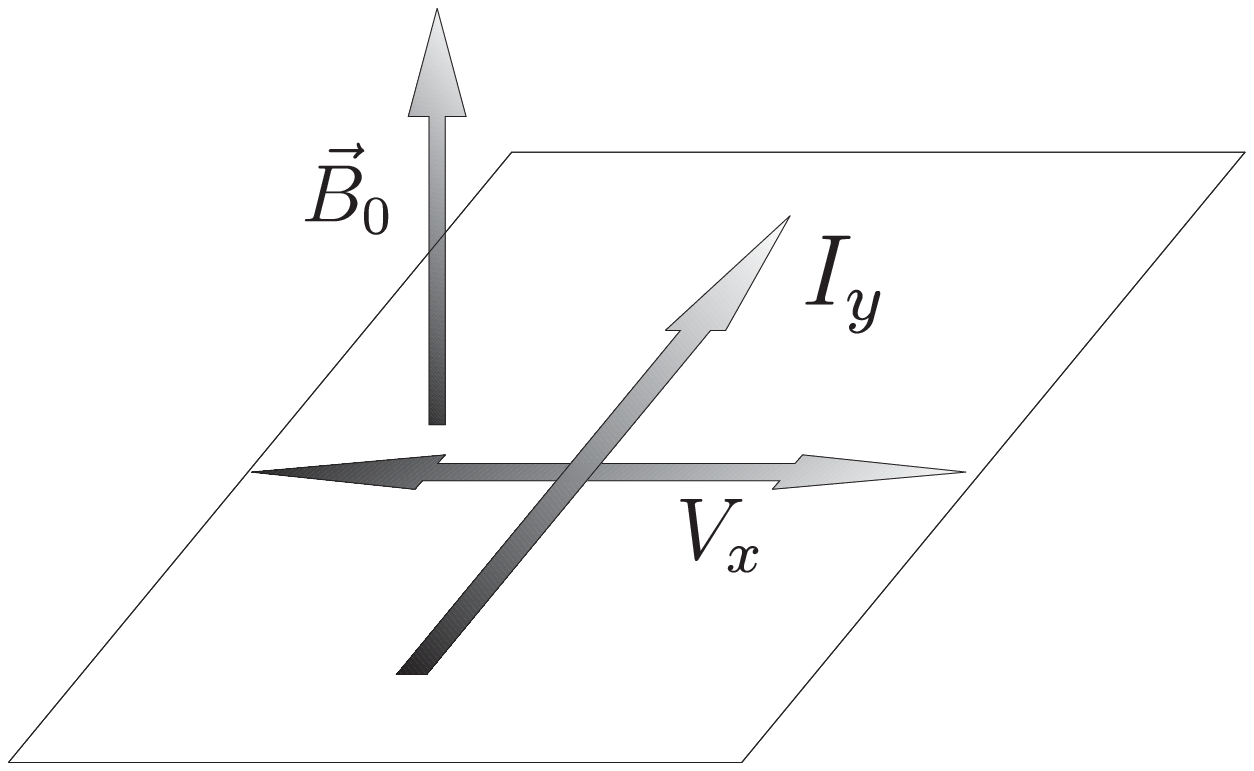}
\caption{Schematic representation of a quantum Hall sample. A voltage drop $V_x$ perpendicular to the current $I_y$ is observed.
\label{fig:sample}
}
\end{figure}

Modern quantum Hall devices are realized in Gallium-Arsenide heterostructures. The electrons are confined to the two-dimensional interface between a layer of doped Al$_{x}$Ga$_{1-x}$As and undoped GaAs. The doped layer is a semi-conductor, while the undoped one is an insulator. By applying a confining electric field perpendicular to the interface (gate voltage), a 2DEG is formed at the interface. In order for an incompressible (Hall) state of the 2DEG to emerge, the device is brought into a strong magnetic field transversal to the interface. A voltage drop $V_y$ may be applied inside the interface so as to generate an electric current $I_y$. Due to the Lorentz force acting on the electrons that carry the current, a voltage drop $V_x$ in the direction perpendicular to the current is then observed (see Fig.~\ref{fig:sample}).

Experimentally, one can measure the longitudinal resistance, $R_L$, as well as the transverse Hall resistance, $R_H$:
\begin{equation*}
  R_L = \frac{V_y}{I_y}\, ,\;\;\;
  R_H = -\frac{V_x}{I_y}\, .
\end{equation*}

Let $n$ denote the {\it density of electrons} in the 2DEG, and let
\begin{equation}
  \Phi_0 = \frac{h c}{e }
\end{equation}
be the quantum of magnetic flux. The dimensionless quantity
\begin{equation}\label{eq:nu}
  \nu = n \frac{\Phi_0}{|{\vec{B}}_{0\perp}|}
\end{equation}
is called the {\it filling factor}. The filling factor corresponds to the number of filled Landau levels for a gas of free spinless fermions of charge $-e$. In Eq.~\eqref{eq:nu}, ${\vec{B}}_{0\perp}$ is the component of the external magnetic field perpendicular to the plane of the 2DEG.

\subsection{Classical theory}
We start by studying the classical mechanics of a 2DEG exhibiting the Hall effect. In a steady state, where the electrons in the 2DEG have a constant velocity, the total force on an electron must vanish. Hence
\begin{equation}\label{eq:lorentz}
  \vec{F}_{e^-\parallel} = -e [\vec{E}_\parallel +\frac{\vec{v}}{c}\wedge \vec{B}_{0\perp}] = 0\, .
\end{equation}
It follows that the velocity of the electrons, $\vec{v}$, is perpendicular to the in-plane electric field $\vec{E}_\parallel$, i.e.,
\begin{equation}
  \vec{E}_\parallel \cdot \vec{v} = 0\, .
\end{equation}
Using \eqref{eq:lorentz}, the electric current density is given by
\begin{equation}
  \vec{j} = -e n \vec{v} = \sigma_H (\vec{e}_z\wedge\vec{E})\, ,
\end{equation}
and the {\it Hall conductivity}, $\sigma_H$, is apparently given by
\begin{equation}\label{eq:hallclassic}
  \sigma_H = R_H^{-1} = \frac{e n c}{|\vec{B}_{0\perp}|} = \frac{e^2}{h}\, \nu\, .
\end{equation}
We observe that classical theory predicts a {\it linear relation} between the Hall conductivity and the filling factor $\nu$, with a factor of proportionality given by $R_K^{-1} = e^2/h$.

\subsection{Experimental behavior of the Hall conductivity}

\begin{figure}[!h]
\center
\includegraphics[width=0.65\textwidth]{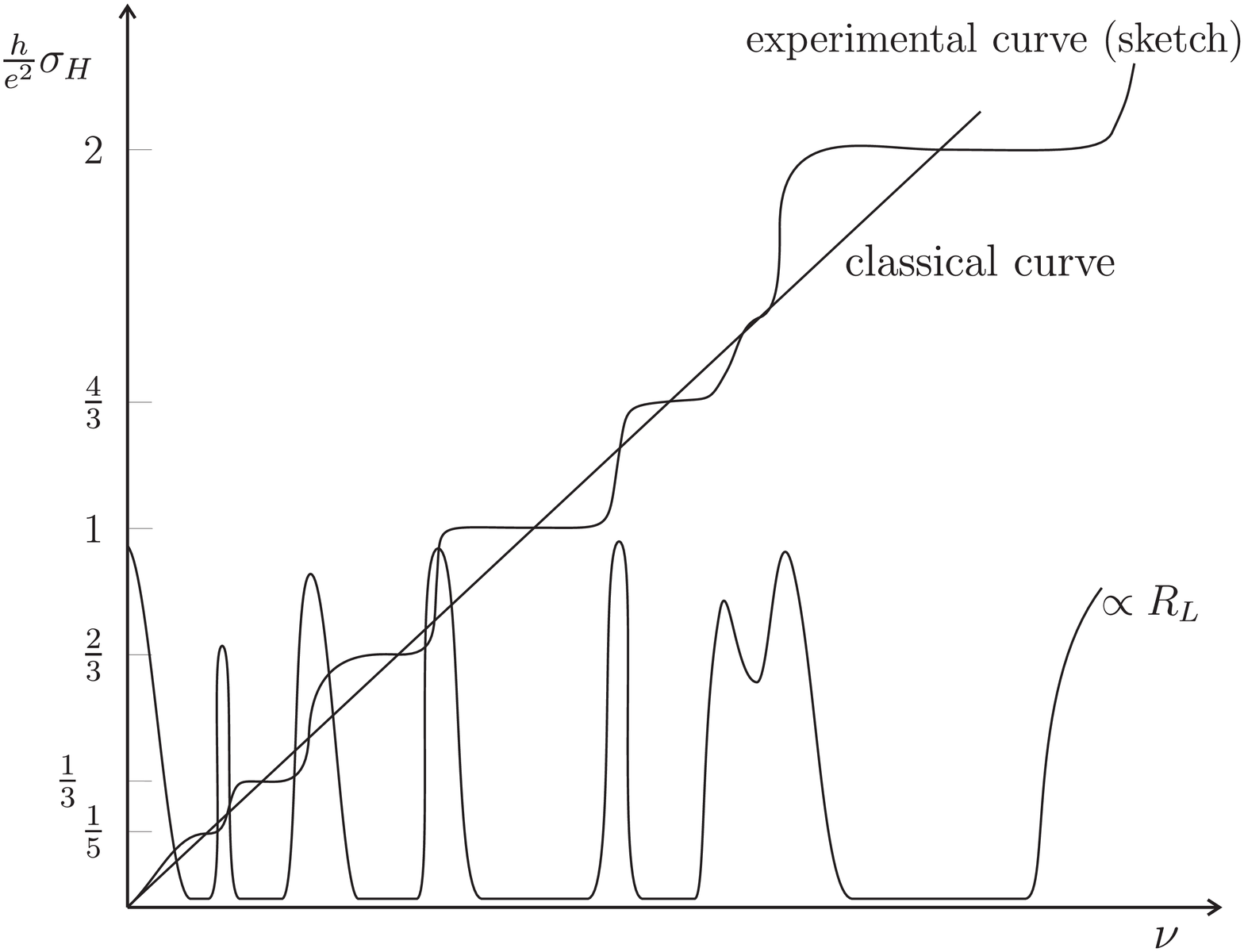}
\caption{Experimental behaviour of the Hall conductivity and the longitudinal resistance of a 2DEG (illustration).
\label{fig:plateaux}
}
\end{figure}

\begin{figure}[!h]
\center
\includegraphics[width=0.87\textwidth]{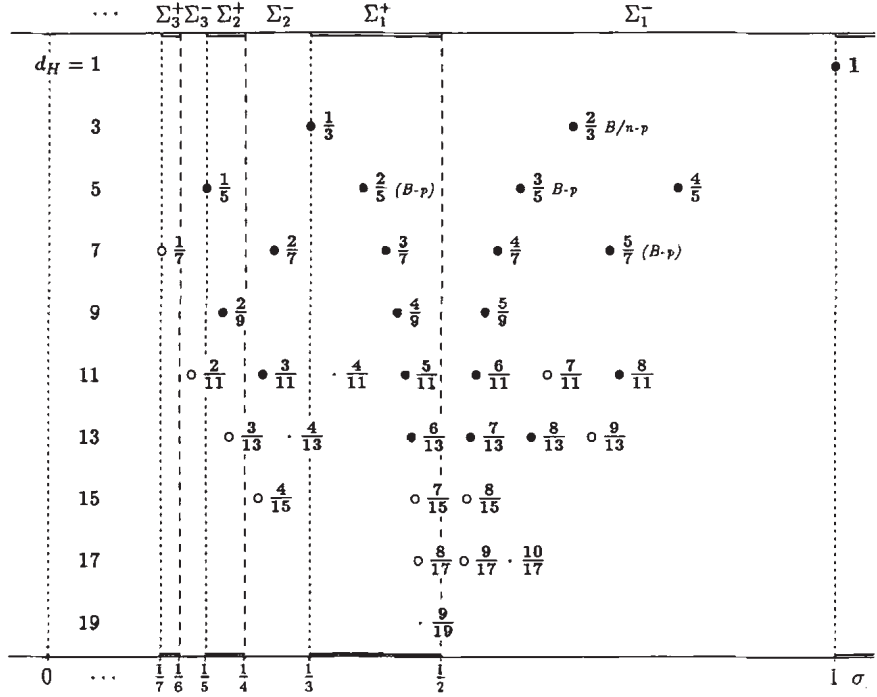}
\caption{Observed Hall plateaux in the range $0 < \sigma \leq 1$; with $\sigma = R_K \sigma_H = \frac{n_H}{d_H}$, where $n_H$ and $d_H$ are co-prime integers.
\label{fig:obseredFrations}
}
\end{figure}

Interestingly, experiments with Hall samples at low temperature and in strong magnetic fields yield a behaviour of $\sigma_H$ that deviates from the classical linear relation in (\ref{eq:hallclassic}). Experimental data, sketched in Fig.~\ref{fig:plateaux}, show plateaux where $\sigma_H$ is very nearly constant. Whenever $(\nu,\sigma_H)$ lies on a plateau, the longitudinal resistivity vanishes and $\sigma_H$ only takes certain values (see Fig.~\ref{fig:obseredFrations}). There is ample experimental evidence for the following claims.
\begin{enumerate}[(I)]
  \item $R_L = 0$ whenever $(\nu,\sigma_H)\in$ plateau \cite{Klitzing80,TsuiStormerGossard82};
  \item plateau heights $\in (e^2/h)\, \mathbbm{Q}$, \cite{Klitzing80,TsuiStormerGossard82};
  \item the cleaner the sample,
  \begin{itemize}
    \item the more plateaux are observed, and
    \item the narrower are the plateaux.
  \end{itemize}
  \item If $R_K \sigma_H \notin \mathbbm{Z}$ ({\it fractional} QHE), some of the quasi-particles observed in the sample appear to carry fractional electric charges \cite{GoldmanSu95,glattli97,Umansky97}.
\end{enumerate}

The precision of the integral plateau heights is of the order of $10^{-9}$. Thus, systems exhibiting the QHE allow for an extremely accurate determination of $R_K = h/e^2$. Together with Josephson junction experiments measuring the fundamental quantity $K_J = e / (h c)$ and quantum pumps, which determine the elementary charge $e$, the {\it metrological triangle} closes \cite{jeanneret2004}.

\subsection{Tasks for theorists}

Given these experimental findings, the following theoretical questions arise:
\begin{enumerate}
  \item
  \begin{enumerate}[(a)]
    \item For what values of $\nu$ is $R_L=0$ (existence of a mobility gap)?
    \item How do the plateau widths scale with disorder?
    \item Quantitative estimates on $|\sigma_H(\nu) - \frac{e^2}{h} \nu|$ ?
    \item Nature of the phase transitions between neighboring incompressible Hall fluids?
    \item Existence of a Wigner crystal for $\nu\lesssim\frac{1}{7}$ ?
  \end{enumerate}
  Answers to these questions would have to be based on a detailed understanding of the quantum many-body problem in the presence of disorder and interactions. In situations relevant for the fractional QHE, quantitative insights are primarily based on large-scale computer simulations \cite{Morf86,Greiter92,Morf02}; but see \cite{Laughlin81,Laughlin83,Halperin82,Haldane83}.
  However, for a 2DEG consisting of {\it non-interacting} electrons in a random external potential, one only observes the integral QHE, and the theory of this phenomenon is well understood \cite{BelissardElstBaldes94,AvronSeilerSimon90}.

  \item
  Assuming that $R_L=0$ (i.e., the 2DEG forms an incompressible fluid), what can we say about
  \begin{enumerate}[(a)]
    \item possible values of $\sigma_H$?
    \item spectrum and properties of quasi-particles?
    \item new experimental tests of theoretical predictions (e.g., interferometry)?
  \end{enumerate}
  Questions of this sort can be studied and answered with the help of an elegant {\it effective field-theory approach}. In the following, we outline this approach.
\end{enumerate}

\subsection{Applications}

The QHE has many important (or potentially important) applications, such as:
\begin{itemize}
  \item Metrology, determination of fundamental constants of nature, definition of a resistance standard, \cite{Klitzing80}.
  \item Novel computer memories.
  \item Q-bits for topological quantum computers (exploitation of quasi-particles with braid statistics) \cite{JF-lectures,kitaev,freedman}.
\end{itemize}

\section{Electrodynamics of an incompressible Hall fluid}

Consider a 2DEG confined to a planar region $\Omega$ and subject to a strong, uniform external magnetic field $\vec{B}_0$ transversal to $\Omega$. In such a system, the vanishing of the longitudinal resistance $R_L$ is a signal for the existence of a {\it mobility gap} in the bulk. One then speaks of an {\it incompressible Hall fluid}. Let us consider the response of the system to a small, slowly time-dependent perturbation of the electromagnetic (EM) fields, with
\begin{equation}
  \vec{B}^{total} = \vec{B}_0 + \vec{B}(x)\,.
\end{equation}
The orbital dynamics of electrons in the region $\Omega$ (assumed to be contained in the $x$-$y$ plane) only depends on $B^{total}_3 = (\vec{B}_0 + \vec{B}(x))\cdot \vec{e}_z$ and $\vec{E}_\parallel = {\bm E}(x) = (E_1(x), E_2(x))$. We set $B = \vec{B}\cdot\vec{e}_z$ and introduce a vector potential,
\begin{equation}
  (A_\mu) := (A_0, A_1, A_2)\,,
\end{equation}
for the electromagnetic field tensor in 2+1 dimensions,
\begin{equation}
  (F_{\mu\nu}) :=
  \left( {
\begin{array}{ccc}
  0    & E_1 & E_2\\
  -E_1 & 0   & -B \\
  -E_2 & B   & 0
\end{array}
  } \right)\, .
\end{equation}
The expectation value of operators in a (quasi-stationary) state of the 2DEG in an external vector potential $A$ is denoted by $\langle (\cdot) \rangle_A$. For example, the electric charge- and current density is given by
\begin{equation}
  j^\mu(x) := \langle \mathcal{J}^\mu(x)\rangle_A\,,
\end{equation}
with $\mu = 0, 1, 2$, where $\mathcal{J}^\mu(x)$ is the quantum-mechanical current density.

From phenomenological and fundamental laws of physics the following equations can be derived:
\begin{enumerate}[(i)]
  \item {\bf Hall's law} (for $R_L=0$)\\
  The electric current is perpendicular to the electric field, i.e.,
\begin{equation}
  j^k(x) = \sigma_H \epsilon^{kl}E_l(x)
\end{equation}
with $k,l=1,2$, where $\epsilon^{kl}$ is the sign of the permutation $(kl)$ of $(12)$, and
\begin{equation}
  x = (x^\mu) = (t,{\bm x})\in \Lambda := \mathbbm{R}\times\Omega\, .
\end{equation}

\item {\bf Charge conservation}\\
Charge- and current density in $\Lambda$ satisfy the continuity equation
\begin{equation}
  \frac{\partial}{\partial t}\, \rho(x) + {\bm \nabla}\cdot{\bm j}(x) = 0\, .
\end{equation}

\item {\bf Faraday's induction law}\\
\begin{equation}
  \frac{\partial}{\partial t}\, B^{total}_3(x) + {\bm \nabla}\wedge{\bm E}(x) = 0\, .
\end{equation}

\end{enumerate}

The laws (i) through (iii) imply that
\begin{equation}\label{eq:diffB}
  \frac{\partial}{\partial t}\, \rho \overset{(ii)} = - {\bm\nabla}\cdot {\bm j} \overset{(i)} = -\sigma_H {\bm \nabla}\wedge{\bm E} \overset{(iii)} = \sigma_H \frac{\partial}{\partial t}\, B^{total}_3\, .
\end{equation}
We integrate Eq.~\eqref{eq:diffB} in time, with integration constants chosen such that
\begin{equation}\begin{split}
  j^0(x) &= \rho(x) + e n\,,\\
  B^{total}_3(x) &= B(x) + B_0\,,
\end{split}
\end{equation}
where $-en$ is the charge density of a homogenous 2DEG in a constant magnetic field $\vec{B}_0$. We then arrive at
\begin{enumerate}[(i)]
\setcounter{enumi}{3}
  \item ``{\bf Chern-Simons Gauss law}" \cite{DeserJackiwTempleton82}\\
\begin{equation}
  j^0(x) = \sigma_H B(x)\, .
\end{equation}
\end{enumerate}

Next, we propose to show that the laws (i) through (iv) imply the existence of anomalous chiral currents circulating at edges of the incompressible Hall fluid.
Faraday's induction law (iii) says that
\begin{equation}\label{eq:faraday}
  \partial_{[\mu} F_{\nu\lambda]} = 0\,,
\end{equation}
which (by Poincar\'e's lemma) implies that the EM field tensor can be derived from a vector potential,
\begin{equation}\label{eq:empotential}
  F_{\mu\nu} = \partial_{[\mu} A_{\nu]}\, .
\end{equation}
In compact notation, laws (i) and (iv) can be written as
\begin{equation}\label{eq:hallcurrent}
  j^\mu(x) = \frac{\sigma_H}{2}\, \epsilon^{\mu\nu\lambda} F_{\nu\lambda}(x) \overset{\eqref{eq:empotential}} = \sigma_H\, \epsilon^{\mu\nu\lambda} \partial_\nu A_{\lambda}(x)\,.
\end{equation}
Whenever $\sigma_H$ is constant, the current \eqref{eq:hallcurrent} satisfies the continuity equation (ii), i.e.,
\begin{equation}
  \partial_\mu j^\mu = \frac{1}{2} \sigma_H \epsilon^{\mu\nu\lambda} \partial_\mu F_{\nu\lambda} \overset{\eqref{eq:faraday}} = 0\, .
\end{equation}
However, wherever the value of $\sigma_H$ {\it jumps}, e.g., at the boundary of the sample, the current \eqref{eq:hallcurrent} is not conserved. Let
\begin{equation}
  \Sigma := support( {\bm \nabla}\sigma_H )\, .
\end{equation}
Then we have that
\begin{equation}\label{eq:noncons}
  \partial_\mu j^\mu(x) = \frac{1}{2} \epsilon^{\mu\nu\lambda} (\partial_\mu\sigma_H)\, F_{\nu\lambda} \neq 0\,, \text{ for } x \in \Sigma\,,
\end{equation}
which violates the law (ii)!

The apparent contradiction between \eqref{eq:noncons} and the continuity equation (ii) disappears when one notices that the current \eqref{eq:hallcurrent} is \emph{not} the total current. Apparently, there must be an additional current supported on $\Sigma$ that cancels the anomaly \eqref{eq:noncons}:
\begin{equation}
  j^\mu = j^\mu_{bulk} \neq j^\mu_{total} = j^\mu_{bulk} + j^{\mu}_{edge}\, ,
\end{equation}
with
\begin{eqnarray*}
  \partial_\mu j^\mu_{total} &=& 0\, ,\\
  support(j^\mu_{edge}) &=& \Sigma\, ,\\
  {\bm j}_{edge}\cdot {\bm \nabla}\sigma_H &=& 0\, .
\end{eqnarray*}
Equation \eqref{eq:noncons} for the {\it bulk current} \eqref{eq:hallcurrent} then implies that, on the ``edge" $\Sigma$,
\begin{equation}\label{eq:anomaly}
  \partial_\mu j^\mu_{edge} = -\partial_\mu j^\mu_{bulk} = \Delta\sigma_H E_{\parallel}|_{\Sigma}\,,
\end{equation}
where $E_{\parallel}|_{\Sigma}$ denotes the electric field ``parallel" to $\Sigma$ (i.e., the component of ${\bm E}|_{\Sigma}$ parallel to the contour lines of $\sigma_H$) and $\Delta\sigma_H$ is the discontinuity of $\sigma_H$ across $\Sigma$. This non-conservation of the edge current is called {\it chiral anomaly} in $1+1$ dimensions. The chiral anomaly (in $3+1$ dimensions) is a well-known phenomenon in gauge theories of elementary particles. It plays an important role in various physical processes; see Refs.~\cite{currentAlgebra,FrohlichPedrini2000}.

\begin{figure}[!h]
\center
\includegraphics[width=0.7\textwidth]{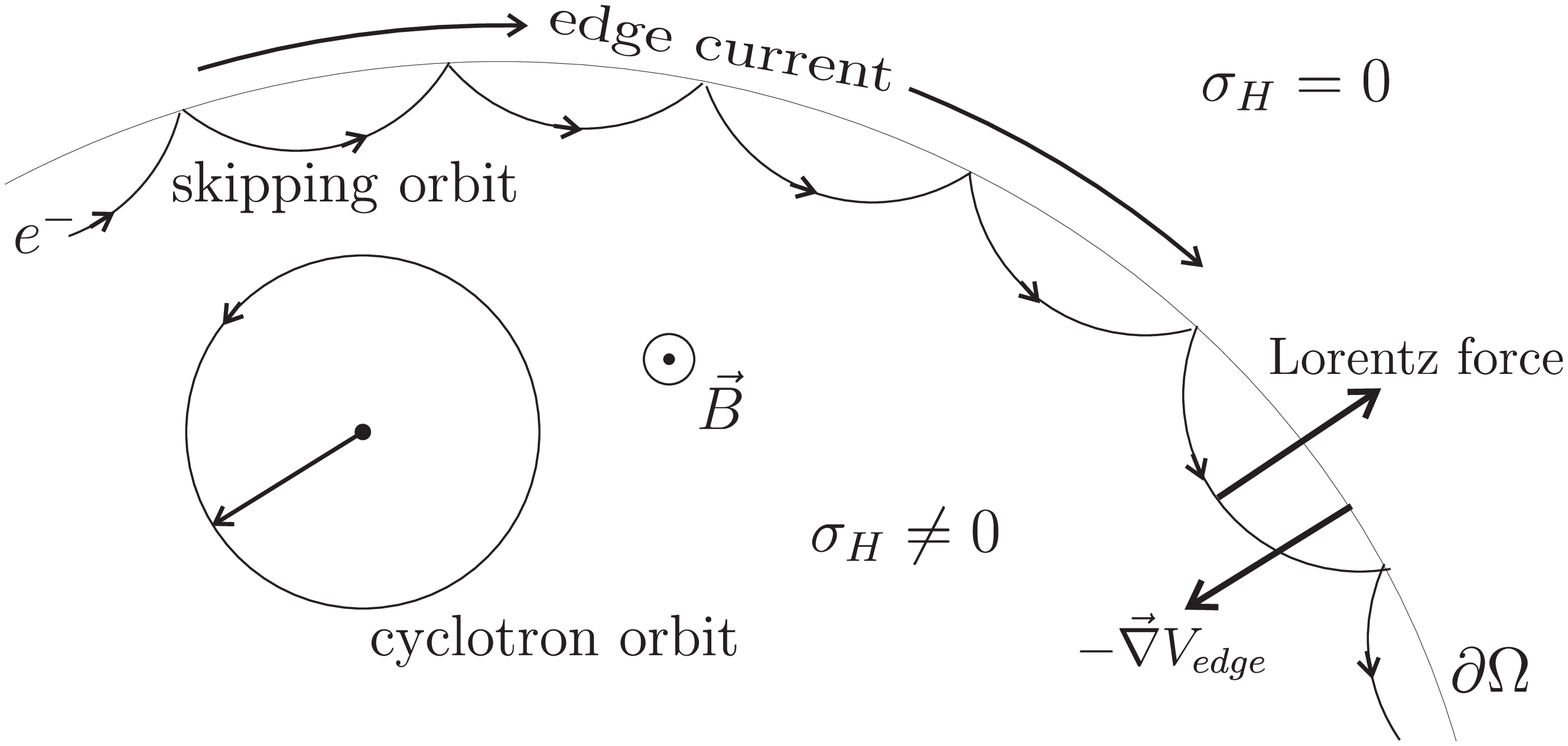}
\caption{Chiral edge current in a 2D electron gas.
\label{fig:hallEdge}
}
\end{figure}
In Fig.~\ref{fig:hallEdge}, an illustration of the edge current in a quantum Hall sample is given. The velocity $\vec{v}$ of an electron at the edge can be calculated by equating the Lorentz force and the confining force,
\begin{equation}
  - e \frac{\vec{v}}{c}\wedge \vec{B} = -\vec{\nabla} V_{edge}\, .
\end{equation}

In classical physics, a phenomenon analogous to the chiral edge currents in an incompressible Hall fluid are the {\it hurricanes} in the atmosphere of the earth! In this case, the magnetic field $\vec{B}$ is replaced by the angular velocity of the earth, $\vec{\omega}_{earth}$, and the role of the Lorentz force is played by the {\it Coriolis force}. The confining force, $-\vec{\nabla} V_{edge}$, in a Hall fluid is replaced by the gradient of the air pressure, $-\vec{\nabla} P$.

\subsection{Chiral anomaly in 1 + 1 dimensions}

An anomalous current satisfying \eqref{eq:anomaly} is carried by charged, chiral, {\it gapless} ``modes", i.e., by particles traveling with a certain velocity along the edge. Let us suppose that the current $j_{edge}^{\mu}$ is carried by $N$ species of chiral modes. We denote their coupling constants to the EM field by $e Q_1,\ldots, e Q_N$. The anomaly of $j^\mu_{edge}$ is then described by (see our discussion in Sect.~\ref{chap:edge})
\begin{equation}
  \partial_\mu j_{edge}^{\mu} = \frac{e^2}{h} (\sum_{i=1}^N Q_i^2)\, E_{\parallel}|_\Sigma\, .
\end{equation}
Combining this equation with \eqref{eq:anomaly} (with $\Delta\sigma_H = \sigma_H$) it follows that the (dimensionless) Hall conductivity is given by
\begin{equation}
  \sigma = R_K \sigma_H = \sum_{i=1}^N Q_i^2\, .
\end{equation}

One can convince oneself that, in the integral QHE, each filled Landau level gives rise to exactly one species of electrons circulating at the edge and thereby contributing to the edge current $j^\mu_{edge}$. Therefore, $R_K \sigma_H = N$ is the number of filled Landau levels. For an incompressible Hall fluid exhibiting the {\it fractional} QHE, with $R_K \sigma_H \notin \mathbb{Z}$, it follows that at least one of the ``charges" $e Q_i$ must be a {\it fraction} of the elementary charge $e$. Arguments similar to the ones reported here can be found, e.g., in \cite{FrohlichKerler91,Wen2,FrohlichPedrini2000}.

\section{Effective action of an incompressible Hall fluid and topological field theory}\label{chap:bulk}

In this section, we determine the effective action of an incompressible Hall fluid~(IHF). Here, and in the following section, we express the Hall conductivity in units of $ e^2 / h$, i.e.,
\begin{equation}
  \sigma := R_K \sigma_H\,,
\end{equation}
where $\sigma$ is {\it dimensionless}. The space-time of the sample is the cylinder $\Lambda = \mathbb{R}\times\Omega$. {\it For simplicity}, we assume that the support of ${\bm\nabla}\sigma$ is $\Sigma = \partial\Omega$; (of course, this is an idealization of what one encounters in real samples). We denote the surface of the cylinder by $\partial\Lambda = \mathbb{R}\times\partial\Omega$. The quantum-mechanical current operator is $\mathcal{J}^{\mu}(x)$, and $\langle(\cdot)\rangle_A$ is the expectation value in a stationary state of the IHF in an external EM field with vector potential $A$.

The {\it effective action} of an IHF, $\Gamma_\Lambda[A]$, is the generating functional of the current Green functions. Hence it satisfies
\begin{equation}\label{eq:currentaction}
  j^\mu_{total}(x) = \langle \mathcal{J}^\mu(x)\rangle_A = \frac{\delta \Gamma_\Lambda[A]}{\delta A_\mu(x)}\, .
\end{equation}
The current in the {\it bulk} of the Hall sample, Eq.~\eqref{eq:hallcurrent}, is given by
\begin{equation}\label{eq:hallcurrent2}
  j^\mu_{total}(x) = j^\mu_{bulk}(x) = \sigma\, \epsilon^{\mu\nu\lambda} \partial_\nu A_{\lambda}(x), \text{  for } x\notin \partial\Lambda\, .
\end{equation}
After integration, \eqref{eq:currentaction} and \eqref{eq:hallcurrent2} then yield the following expression for the effective action.
\begin{equation}\label{eq:effeA}
  \Gamma_\Lambda[A] = \frac{\sigma}{2}\int_\Lambda d^3x\, \epsilon^{\mu\nu\lambda} A_\mu(x) \partial_\nu A_{\lambda}(x) + \frac{1}{2}\, \Gamma[a] = \frac{\sigma}{2}\int_\Lambda A \wedge dA + \frac{1}{2}\, \Gamma[a]\, ,
\end{equation}
where $a := A|_{\partial\Lambda}$, and $\Gamma[a]$ is the generating functional of the edge-current Green functions (up to terms local in $a$).

The {\it Chern-Simons action} $\int_\Lambda A \wedge dA$ in \eqref{eq:effeA} is \emph{not invariant} under a gauge transformation, $A\mapsto A + d\alpha$ with $\alpha|_{\partial\Lambda} \neq 0$. In fact, we find that
\begin{equation}\label{eq:gaugeanomaly}
  \delta\int_\Lambda A \wedge dA = \int_\Lambda d\alpha\wedge dA = \int_{\partial\Lambda} \alpha\, da\,,
\end{equation}
where we have used Stokes' theorem.
In our case of a time-independent sample $\Omega$, we have that $da = \partial_\mu a_\nu\, dx^\mu\wedge dx^\nu = E_{\parallel}\, dt\wedge d\xi$, where $\xi$ is a coordinate parametrizing $\partial\Omega$, and
\begin{equation}
  E_{\parallel} = \epsilon^{\mu\nu}\partial_\mu a_\nu
\end{equation}
is the electric field parallel to the edge. The total action $\Gamma_\Lambda$, however, {\it must be gauge invariant} (conservation of electric charge). Therefore, the violation of gauge invariance described in \eqref{eq:gaugeanomaly} must be cancelled by the edge action $\Gamma[a]$, which is then found to be given by
\begin{equation}\label{eq:effa}
  \Gamma[a] = \frac{\sigma}{2} \int_{\partial\Lambda} d^2\xi\, \{( E_{\parallel} - \partial_\mu a^\mu) \Box^{-1} ( E_{\parallel} - \partial_\nu a^\nu) + a_\mu a^\mu \}
\end{equation}
(up to manifestly gauge-invariant terms), assuming that all chiral edge modes have the same propagation speed and direction; (the general case will be discussed in Sect.~\ref{chap:edge}). By \eqref{eq:currentaction} and \eqref{eq:effeA},
\begin{equation}\label{eq:edgeeffcurrent}
  j^\mu_{edge} = \frac{1}{2} \{ \sigma\, \epsilon^{\mu\nu} a_\nu + \frac{\delta\Gamma[a]}{\delta a_\mu} \} \,,
\end{equation}
where $\mu, \nu \in \{0,1\}$, and, using \eqref{eq:effa},
\begin{equation}\label{eq:edgeeffdivergence}
  \partial_\mu j^\mu_{edge} = \sigma\, \epsilon^{\mu\nu}\partial_\mu a_\nu\,,
\end{equation}
in accordance with Eq.~\eqref{eq:anomaly}.

We would like to emphasize that, up to gauge invariant terms, the effective action for the edge current, \eqref{eq:effa}, is uniquely determined by the requirement of electric charge conservation. The only possible generalization is to consider several independent edge channels of charged quasi-particles. In contrast to the bulk contribution to the total action \eqref{eq:effeA}, the edge action is {\it not} topological, i.e., it depends on the {\it space-time metric} of the edge. Therefore, in general, each edge channel may couple to a different space-time metric (i.e., exhibit a different propagation speed). We will discuss this point in more detail in the next section.

The total electric current, $\mathcal{J}^\mu$, is conserved,
\begin{equation}
  \partial_\mu \mathcal{J}^\mu = 0\, .
\end{equation}
By Poincar\'e's lemma, it can therefore be derived from a vector potential, which we denote by $B$,
\begin{equation}\label{eq:bulkcurrent}
  \mathcal{J}^\mu = \sqrt{\sigma}\, \epsilon^{\mu\nu\lambda}\, \partial_\nu B_\lambda\, .
\end{equation}
The potential $B$ in \eqref{eq:bulkcurrent} gives rise to the gauge symmetry $B_\mu\mapsto B_\mu + \partial_\mu \beta$: $\mathcal{J}^\mu$ does {\it not} change under a gauge transformation of $B$. The action
\begin{equation}\label{eq:effAB}
  S_\Lambda[B,A] = \frac{1}{2} \int_\Lambda B\wedge dB + \int_\Lambda d^3x\, \mathcal{J}^\mu A_\mu + \tilde S[B|_{\partial\Lambda},a]
\end{equation}
describes the theory of the gauge potential $B$ coupled to an EM vector potential $A$. 
In \eqref{eq:effAB}, $\tilde S$ is the edge action that makes the total action gauge-invariant.
With an appropriate choice of $\tilde S$, the action \eqref{eq:effAB} yields the effective action \eqref{eq:effeA}, after functional integration over the field $B$.

$S_\Lambda[B,A]$ is the action of a topological U(1) Chern-Simons theory. The charge operator associated with a region $\mathcal{O}$ of $\Omega$ is defined as
\begin{equation}
  Q_\mathcal{O} := \int_\mathcal{O} d^2x\, \mathcal{J}^0(t,{\bm x}) = \sqrt{\sigma}\, \int_{\partial \mathcal{O}} B\, .
\end{equation}
Thus, the exponential of $Q_\mathcal{O}$,
\begin{equation}
  e^{i Q_\mathcal{O}} = e^{i \sqrt{\sigma} \int_{\partial \mathcal{O}} B }\,,
\end{equation}
is a {\it Wilson loop operator} for the field $B$ associated with the contour $\partial \mathcal{O}$. Wilson loops and -networks furnish the ``observables" in a 3D topological field theory (TFT). Static sources of $B$ inserted in the bulk at a point $z\in\Omega$ are described by vectors in a Hilbert space,
\begin{equation}
  |(q,\lambda), z \rangle \in   [(q,\lambda), z ]
\end{equation}
with
\begin{equation}\label{eq:charge}
  Q_\mathcal{O} |(q,\lambda), z \rangle = \sqrt{\sigma}\, q\, |(q,\lambda), z\rangle\,,
\end{equation}
whenever $\mathcal{O}$ contains the insertion point $z$. Here, $q$ is the flux of the field $B$, and $\lambda$ is some {\it additional ``internal" quantum number} needed to label the sectors of the TFT describing the bulk of an IHF. The state vectors $|(q,\lambda),z\rangle$ are elements of a sector (subspace) of the total state space denoted by
\begin{equation}
  [(q,\lambda), z]\,.
\end{equation}
Sectors are thus labeled by $(q,\lambda)$ and an insertion point $z\in\Omega$. The fact that the bulk theory has trivial dynamics (static sources, purely local current correlators) is a consequence of the mobility gap in the bulk, after passing to the scaling limit.

\subsection{Fusion of sources}

\begin{figure}[!h]
\center
\includegraphics[width=0.4\textwidth]{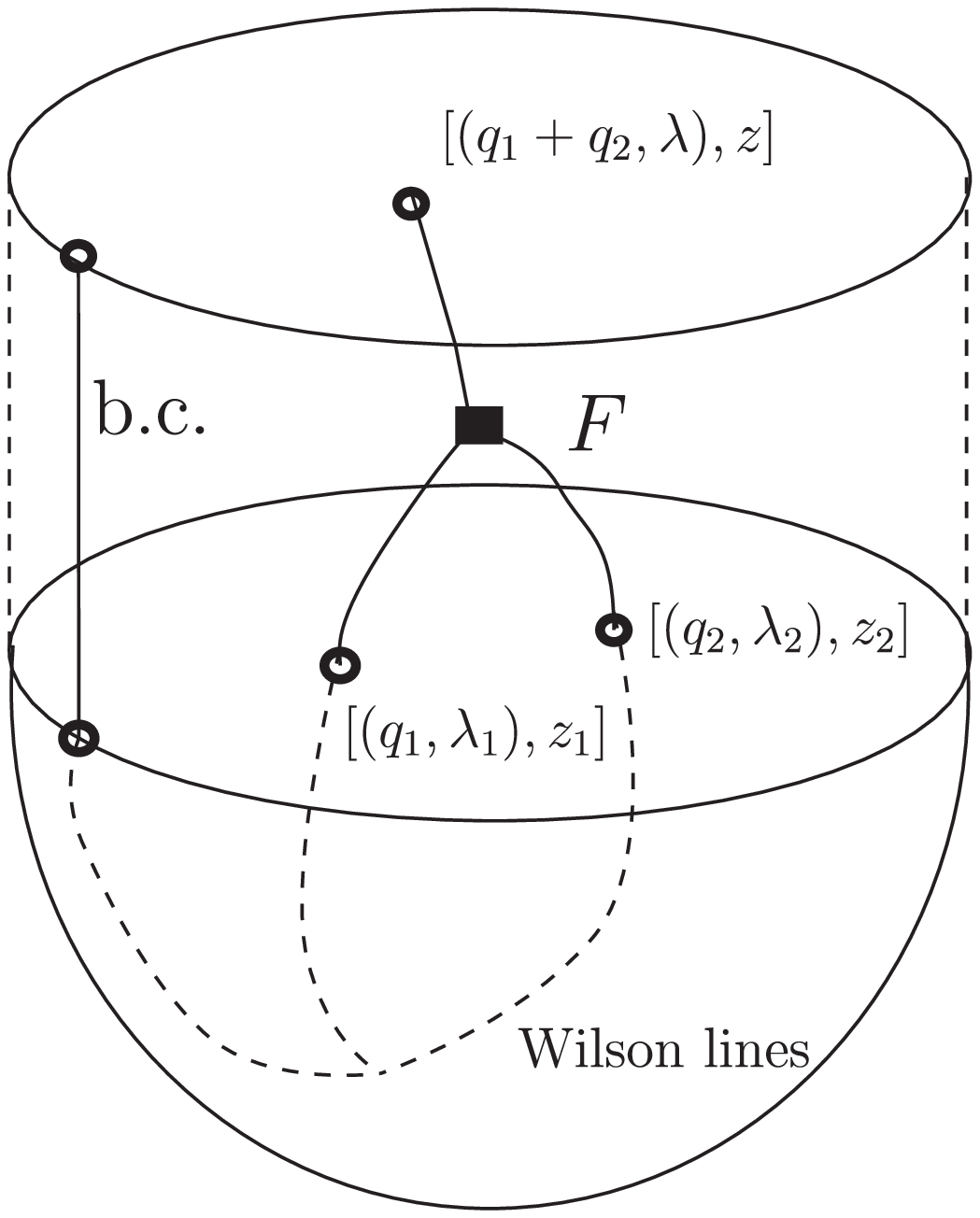}
\caption{Fusion of two sources in a topological field theory; (``b.c." stands for a boundary condition).
\label{fig:fusion}
}
\end{figure}

Next, we discuss properties of states in a TFT with several distinct sources. The sources in a TFT constitute a {\it fusion algebra}. This means the following: Consider the tensor product space corresponding to two sources located at $z_1$ and $z_2$, denoted by $[(q_1,\lambda_1), z_1]\otimes[(q_2,\lambda_2), z_2]$. As the locations $z_1$ and $z_2$ of the two sources approach the same point $z$, the tensor product can be written as a direct sum,
\begin{equation}\label{eq:fusionspace}
  [(q_1,\lambda_1), z_1] \otimes [(q_2,\lambda_2), z_2] \simeq \underset{\lambda}\oplus\, [(q,\lambda), z]\otimes 
  \mathbb{C}^{N^\gamma_{\gamma_1\gamma_2}}\,,
\end{equation}
where $\gamma_i = (q_i,\lambda_i)$ and $q=q_1+q_2$. The non-negative integers $N^\gamma_{\gamma_1\gamma_2}$ are called {\it fusion rules}; they are the multiplicities of the spaces $[(q,\lambda), z]$ in the tensor product space. The morphisms, $F^{\gamma, a}_{\gamma_1\gamma_2}$, from $[\gamma_1, z_1] \otimes [\gamma_2, z_2]$ to the space $[\gamma, z]$ are called {\it fusion matrices},
\begin{equation}\label{eq:fusionmorphism}
  F^{\gamma, a}_{\gamma_1\gamma_2}: [(q_1,\lambda_1), z_1 ] \otimes [(q_2,\lambda_2), z_2 ] \rightarrow [(q,\lambda), z ]_a ,
\end{equation}
with $a=1,\ldots,N^\gamma_{\gamma_1\gamma_2}$; (see Fig.~\ref{fig:fusion}).

In a ``physical" theory, i.e., for a {\it quasi-rational TFT},  all multiplicities in the ``Clebsch-Gordan series" \eqref{eq:fusionspace} must be finite, more precisely, $\sum_\gamma N^\gamma_{\gamma_1\gamma_2}<\infty$, for all pairs $\{\gamma_1,\gamma_2\}$. If $N^\gamma_{\gamma_1\gamma_2}> 0$, then $\gamma$ is called a {\it fusion channel} for $\gamma_1$ and $\gamma_2$.
A TFT is {\it abelian} if $\sum_\gamma N^\gamma_{\gamma_1\gamma_2} = 1$, for all pairs $\{\gamma_1,\gamma_2\}$. In Sect.~\ref{chap:edge}, we will focus on edge theories dual to abelian TFTs in the bulk.

\subsection{Constraints on physical TFTs}\label{sec:spin}

The spin $s_{q,\lambda}$ of a state is determined by considering a rotation in the plane through an angle of $2\pi$. Let $U_{rot}(2\pi)$ represent the rotation by $2\pi$ around the origin. Then
\begin{equation}\label{eq:rotation}
  U_{rot}(2\pi)|(q,\lambda), z \rangle = e^{2\pi i s_{q,\lambda}} |(q,\lambda), z \rangle \, ,
\end{equation}
and the spin $s_{q,\lambda}$ is given by (see, e.g., \cite{FrohlichPedriniSchweigertWalcher01,FrohlichKing89})
\begin{equation}\label{eq:spin}
  s_{q,\lambda} = \frac{q^2}{2} + \Delta_\lambda\, ,
\end{equation}
where, for a quasi-rational TFT, as defined above, \emph{Vafa's theorem} \cite{vafa88} implies that
\begin{equation}
  \Delta_\lambda\in\mathbbm{Q}\, .
\end{equation}
So, in general, $s_{q,\lambda} \notin \frac{1}{2}\mathbbm{Z}$, and the theory may have quasi-particles in its spectrum, that have fractional spin and are neither bosons nor fermions (so-called {\it anyons}).

In a theory describing a physical IHF, there must exist bulk states with the quantum numbers and properties of one-electron states. Suppose that the state $|(q^*\!,\lambda^*), z\rangle$ is obtained by adding a single electron at the point $z\in\Omega$ to the groundstate of the IHF. Then we have the following constraints: The charge of this state, see \eqref{eq:charge}, must be
\begin{equation}
  \sqrt{\sigma}\, q^* = -1\, .
\end{equation}
Furthermore, the spin, see \eqref{eq:spin}, must be half-integer, i.e.,
\begin{equation}
  s_{q^*\!, \lambda^*} = \frac{(q^*)^2}{2} + \Delta_{\lambda^*} = \frac{1}{2\sigma} + \Delta_{\lambda^*} = l + \frac{1}{2}
\end{equation}
with $l\in\mathbbm{Z}$. From Vafa's theorem we know that $\Delta_{\lambda^*}$ is rational. \emph{It thus follows that the Hall conductivity}, $\sigma$, \emph{is rational},
\begin{equation}
  \sigma = \frac{n_H}{d_H} \in \mathbbm{Q}\, .
\end{equation}

There is a third constraint on physical theories: the so-called {\it relative locality} of all quasi-particle states with respect to electron insertions. We will not discuss it here; but see Refs.~\cite{FrohlichThiran94,FrohlichPedriniSchweigertWalcher01} for more information. Using these three constraints on a theory describing a physical IHF, one can show that the smallest electric charge of a quasi-particle that can appear in an IHF is given by
\begin{equation}
  q_{min} = \frac{e}{f\, d_H}\,,
\end{equation}
where $f \in\mathbbm{N}$ is an integer (namely the order of the simple current corresponding to the insertion of an electron); see, e.g., \cite{FrohlichPedriniSchweigertWalcher01}.

To conclude this section, we remark that one may view the spaces labeled by $(q,\lambda)$ as sectors of a chiral algebra describing some chiral conformal field theory (CFT) \cite{Difrancesco97}. Abstractly, they can be understood as the ``irreducible objects" of a braided tensor category \cite{tensorCategories,FuchsRunkelSchweigert}.

\subsection{Monodromy and braiding}\label{sec:monodromy}

\begin{figure}[!h]
\center
\includegraphics[width=0.5\textwidth]{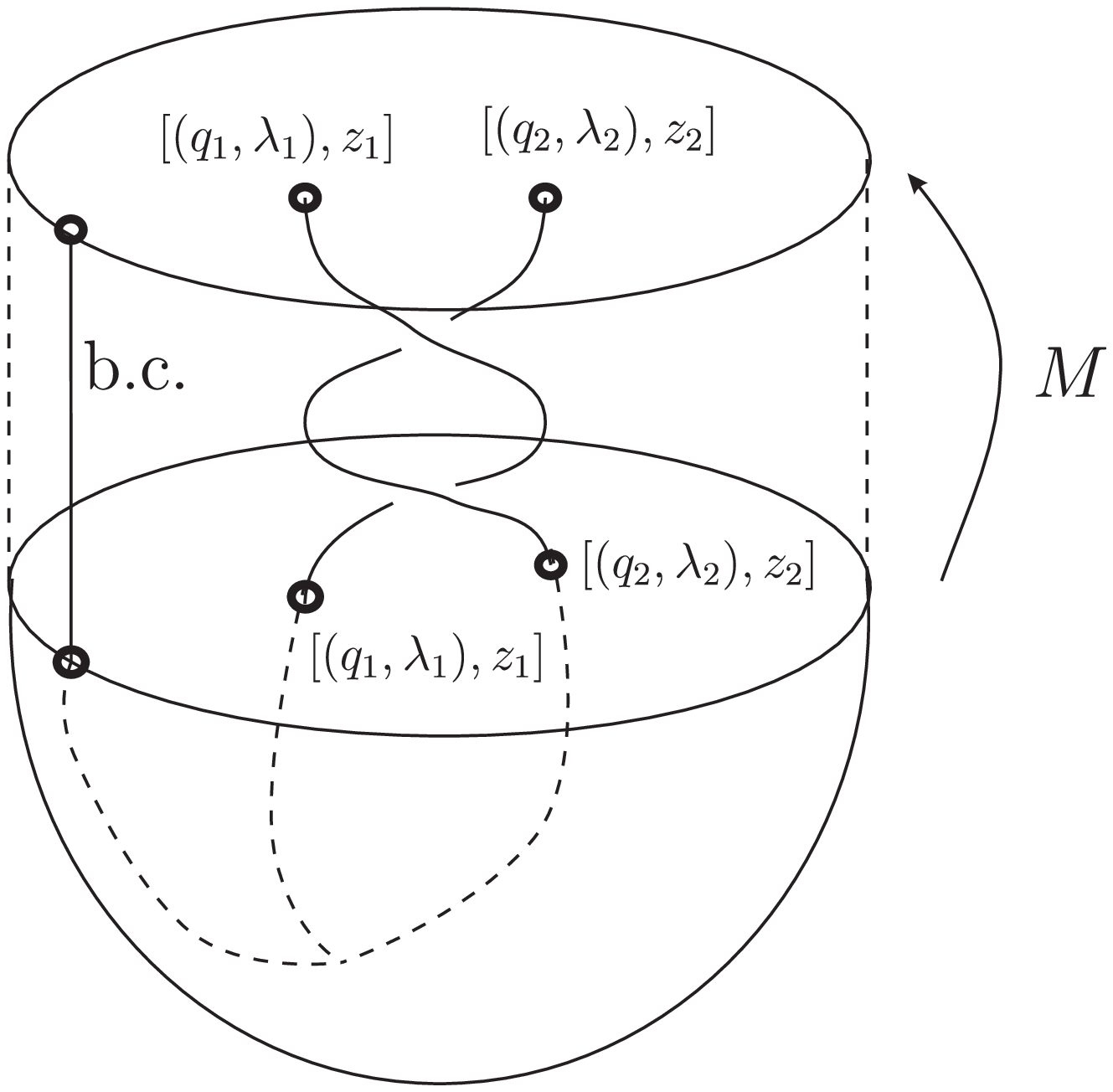}
\caption{Monodromy operation for a pair of particles with quantum numbers $(q_1,\lambda_1)$ and $(q_2,\lambda_2)$.
\label{fig:monodromy}
}
\end{figure}

Let us consider the transformation of a state describing two sources when the sources are adiabatically carried around one another, as depicted in Fig.~\ref{fig:monodromy}. This corresponds to a rotation of the two sources through an angle $2\pi$. After subtracting the contribution of the spins of the sources, the {\it monodromy matrix}, $M$, is defined by
\begin{equation}
  U_{rot}(2\pi)|_{[(q_1,\lambda_1), z_1] \otimes [(q_2,\lambda_2), z_2 ]} = e^{2\pi i (s_{q_1\!,\lambda_1} + s_{q_2\!,\lambda_2})} M_{(q_1,\lambda_1)(q_2,\lambda_2)}
\end{equation}
We may fuse the tensor product states on both sides. Using \eqref{eq:fusionmorphism} and \eqref{eq:rotation}, we get
\begin{equation}
  U_{rot}(2\pi)|_{[(q_1,\lambda_1), z_1] \otimes [(q_2,\lambda_2), z_2 ]} = \underset{\lambda}\oplus\, e^{2\pi i s_{q,\lambda} } F^{(q,\lambda),a}_{(q_1,\lambda_1) (q_2,\lambda_2) }\, .
\end{equation}
This shows that the monodromy matrix $M_{(q_1,\lambda_1)(q_2,\lambda_2)}$ is block-diagonal in the decomposition of the tensor product space into the subspaces $[(q,\lambda), z]$, and its eigenvalues on these subspaces are given by
\begin{equation}\label{eq:monodromy}
  e^{2\pi i ( s_{q_1+q_2\!,\lambda} - s_{q_1\!,\lambda_1} - s_{q_2\!,\lambda_2})} = e^{2\pi i\, q_1 q_2}\, e^{2\pi i (\Delta_\lambda -\Delta_{\lambda_1} -\Delta_{\lambda_2})}\, .
\end{equation}
The factor $e^{2\pi i q_1 q_2}$ corresponds to the well-known {\it Aharonov-Bohm phase} for carrying a charged particle around an insertion of magnetic flux.
In general, it may happen that $M\neq \mathbbm{1}$, for some pairs of quasi-particles. The particles then exhibit braid statistics.

Braid statistics is an interesting phenomenon only encountered in two-dimensional systems. In dimensions larger than two, quantum statistics is \emph{always} described by representations of the permutation group; see, e.g., \cite{SpinOrActually07} and refs.\ given there. The theoretical possibility of braid statistics in 2D systems appears to be realized in IHFs at certain fractional plateaux.

\section{The edge of an incompressible Hall fluid\label{chap:edge}}

Next, we propose to find an action, $S$, for matter fields located on the edge, $\partial\Lambda$, that describe chiral modes coupled to the electromagnetic field. (In addition, there may be {\it neutral} modes, which we omit here.) The main constraint on the edge action is its gauge variation; i.e., under
\begin{equation}
  a \mapsto a + d\alpha|_{\partial\Lambda}\, ,
\end{equation}
we find [cf. Eq.~\eqref{eq:gaugeanomaly}] that
\begin{equation}\label{eq:ga}
  \delta S = - \sigma \int_{\partial\Lambda} d^2\xi\, \alpha\, \epsilon^{\mu\nu}\partial_\mu a_\nu\, .
\end{equation}
Furthermore, the edge current, $J_{edge}^\mu$, must satisfy the anomaly equation [see \eqref{eq:edgeeffdivergence}]
\begin{equation}\label{eq:anomaly2}
  \partial_\mu J^\mu_{edge} = \sigma\, \epsilon^{\mu\nu}\partial_\mu a_\nu\, .
\end{equation}

Since the edge degrees of freedom of an IHF form a system in 1+1 dimensions, we can make use of bosonization techniques. A current carried by gapless quasi-particles can be decomposed into left- and right-moving currents with opposite propagation directions, $J_L$ and $J_R$. The vector current, $J = J_L + J_R$, is always conserved. This means that the anomalous edge current, $J_{edge}$, must be {\it chiral}, i.e., there is an imbalance between left- and right-moving modes.

Conservation of the vector current, $\partial_\mu J^\mu  = 0$, allows us to introduce a (possibly multi-valued) scalar potential $\phi$, i.e.,
\begin{equation}
  J^\mu = \epsilon^{\mu\nu} \partial_\nu \phi\, .
\end{equation}
In the {\it absence} of an external electric field, let us write the chiral edge current in terms of the scalar potential as
\begin{equation}
  J^\mu_{edge} = \frac{\sqrt{\sigma}}{2}(\partial^\mu\phi + \epsilon^{\mu\nu}\partial_\nu\phi)\, .
\end{equation}
When the external field vanishes, the edge current is conserved,
\begin{equation}
  \partial_\mu J^\mu_{edge} \propto \Box\,\phi = 0\,,
\end{equation}
which is the equation of motion for a massless free Bose field, $\phi$, with action
\begin{equation}
  S[\phi,a=0] =  \int \sqrt{|g|}\, d^2\xi\, \frac{1}{2} g^{\mu\nu}\partial_\mu\phi\, \partial_{\nu}\phi\, .
\end{equation}
We choose a metric $g^{\mu\nu}$ on $\partial\Lambda$ with $g=\det (g^{\mu\nu}) = -1$. More precisely,
\begin{equation}
  (g^{\mu\nu}) = \text{diag}(u^{-1},-u)\,,
\end{equation}
where $u$ is the propagation speed of $\phi$.

Next, we introduce the edge action, $S[\phi,a]$, for a non-zero vector potential $a_\mu$: $S[\phi,a]$ is required to yield the effective edge action \eqref{eq:effa}, after functional integration over the matter field $\phi$, i.e.,
\begin{equation}
  \int D\phi\, e^{2\pi i\, S[\phi,a]} = e^{2\pi i\, \Gamma[a]}\,.
\end{equation}
This uniquely fixes $S[\phi,a]$ (up to gauge-invariant terms). It is found to be given by
\begin{equation}\label{eq:action2}
  S[\phi,a] =  \int d^2\xi\, \{ \frac{1}{2} \partial^\mu\phi\, \partial_{\mu}\phi + \sqrt{\sigma}\, (\partial^\mu\phi + \epsilon^{\mu\nu}\partial_\nu\phi)\, a_\mu  + \frac{\sigma}{2}a^\mu a_\mu\, \} \,,
\end{equation}
with $\partial^\mu\phi = g^{\mu\alpha}\partial_\alpha\phi$ and $a^\mu = g^{\mu\alpha} a_\alpha$.
The edge current in the {\it presence} of external fields [see \eqref{eq:edgeeffcurrent}] is
\begin{equation}\label{eq:edgecurrent2}
  J^\mu_{edge} = \frac{1}{2}\{ \sigma\, \epsilon^{\mu\nu} a_\nu + \frac{\delta S}{\delta a_{\mu}}\}
   = \frac{1}{2}\{\sqrt{\sigma}\, (\partial^\mu\phi + \epsilon^{\mu\nu}\partial_\nu\phi) +  \sigma(a^\mu + \epsilon^{\mu\nu} a_\nu) \} \,.
\end{equation}
It exhibits the correct anomaly \eqref{eq:anomaly2}, on a solution to the equation of motion for $\phi$.

To generalize our construction, $N>1$ conserved vector currents can be introduced,
\begin{equation}
  J_i^\mu = \epsilon^{\mu\nu} \partial_\nu \phi_i\,.
\end{equation}
The action for the fields ${\bm \phi} = (\phi_1,\ldots,\phi_N)$ generalizing \eqref{eq:action2} is then given by
\begin{equation}\label{eq:action2i}
  S[{\bm\phi},a] = \sum_i \int d^2\xi\, \{ \frac{1}{2}\partial^\mu\phi_i \partial_\mu\phi_i + Q_i\,(\partial^\mu\phi_i + \chi_i\epsilon^{\mu\nu}\partial_\nu\phi_i)\, a_\mu + \frac{Q_i^2}{2}\, g_i^{\mu\nu} a_\mu a_\nu \}\,,
\end{equation}
where $\chi_i\in\{+,-\}$ is the {\it chirality} of the edge current carried by the field $\phi_i$, and $Q_i\in \mathbb{R}$ are some constants. The metrics, $g_i$, may be different for each field, $(g_i^{\mu\nu}) = diag(u_i^{-1},-u_i)$, where $u_i$ is the propagation speed of $J_i$. The action \eqref{eq:action2i} has the correct gauge variation, and the edge current,
\begin{equation}\label{eq:edgei}
  J^\mu_{edge} = \sum_{i=1}^N J^\mu_{\chi i} = \frac{1}{2} \sum_{i=1}^N Q_i\, \{ \partial^\mu\phi_i + \chi_i \epsilon^{\mu\nu}\partial_\nu\phi_i + Q_i ( a^\mu + \chi_i \epsilon^{\mu\nu} a_\nu) \}\,,
\end{equation}
exhibits the expected anomaly, provided that
\begin{equation}\label{eq:sigmai}
  \sum_i \chi_i Q_i^2 = \sigma\, .
\end{equation}

Using \eqref{eq:action2i}, the equation of motion for the field $\phi_i$ is given by
\begin{equation}\label{eq:eqmotion}
  g^{\mu\nu}_i \partial_\mu ( \partial_\nu\phi_i + Q_i a_\nu) = \chi_i Q_i E\,.
\end{equation}
By inspection, under a gauge transformation $a_\mu\mapsto a_\mu + \partial_\mu\alpha$, solutions to \eqref{eq:eqmotion} transform as
\begin{equation}
  \phi_i\mapsto\phi_i - Q_i \alpha\, .
\end{equation}
This shows that the edge currents in \eqref{eq:edgei} [and, in particular, \eqref{eq:edgecurrent2}] are gauge-invariant objects.

Canonical quantization of the action \eqref{eq:action2i} yields the equal-time commutators
\begin{equation}\label{eq:currentalgebra}
  [J_{\chi j}^0(x,t),J_{\chi k}^0(y,t)] = \frac{i}{2\pi} \chi_j\, Q_j^2\, \delta_{jk}\, \delta'(x-y)\, .
\end{equation}
Hence, the currents $J_{\chi i}$ generate $N$ chiral U(1) Kac-Moody algebras \cite{goddardOlive86}. Using \eqref{eq:currentalgebra} and \eqref{eq:sigmai}, the edge current \eqref{eq:edgei} satisfies the commutation relation
\begin{equation}
  [J_{edge}^0(x,t), J_{edge}^0(y,t)] = i \frac{\sigma}{2\pi}\, \delta'(x-y)\,.
\end{equation}

The cancellation of the gauge anomaly of the electromagnetic effective (Chern-Simons) action in the bulk by appropriate massless chiral field theories on the edge of the Hall sample is an example of the {\it holographic principle} (applied, here, to gapless quantum field theories in two dimensions and three-dimensional TFTs). A more conventional version of this principle tells us that there is a correspondence between certain 3D TFTs and 2D chiral conformal field theories (CFTs). This formulation is somewhat misleading, though, since (as our example shows) the massless edge modes may have different propagation speeds, i.e., the conformal symmetry may be broken. This is usually the case in realistic IHFs.

For abelian IHFs with $N$ conserved currents, \eqref{eq:action2i}, the family of physical theories has been classified mathematically, \cite{FrohlichStuder93,FrohlichThiran94,FrohlichStuderThiran95,FrohlichStuderKerlerThiran95}. For each fluid, it is possible to enumerate all quasi-particle excitations. For a given Hall conductivity, $\sigma$, and a certain number, $N$, of currents, the quasi-particles are labeled by vertices of a lattice, $\Gamma^*$, of dimension $N$, dual to an {\it odd integral lattice}, $\Gamma$, of (multi-)electron excitations. This again implies that the dimensionless Hall conductivity $\sigma$ is a {\it rational number}. Steps towards a generalization of this approach to \emph{non-abelian} IHFs have been undertaken in \cite{FrohlichPedriniSchweigertWalcher01,BoyarskyCheianovFrohlich09,MooreRead91,Wen1}.

\subsection{Chiral vertex operators}

So far, we have introduced an algebra of chiral currents, $\{ J^\mu_{\chi j} \}$, generating U(1) Kac-Moody algebras; Eq.~\eqref{eq:currentalgebra}. The next step is to construct quasi-particle creation- and annihilation operators as {\it chiral vertex operators}. For simplicity, let us discuss the case of an abelian IHF with only a {\it single} edge degree of freedom. The operator
\begin{equation}\label{eq:qp}
  \psi_{q}(\xi) = \mathcal{N} \exp [2\pi i\, \frac{q}{\sqrt{\sigma}} \int^{\xi} dy^\mu\, \{ \epsilon_{\mu\nu}J^\nu_{edge} - \sigma a_\mu\} ]\,,
\end{equation} 
creates a charged quasi-particle at a point $\xi = (\xi^\mu) \in\partial\Lambda$ of the edge. $\mathcal{N}$ denotes normal ordering, and $J^\mu_{edge}$ is given in \eqref{eq:edgecurrent2}. The starting point of the line integral in \eqref{eq:qp} is some reference point, usually taken to be an ohmic contact. Note that a continuous deformation of the path in the line integral in \eqref{eq:qp} leaves $\psi_q(\xi)$ invariant. In other words, the vertex operators only depend on the homotopy class of the path. This is because the curl of the integrated vector field vanishes,
\begin{equation}
  \epsilon^{\alpha\mu} \partial_\alpha (\epsilon_{\mu\nu} J^\nu_{edge} - \sigma a_\mu ) = 0\,.
\end{equation}

The electric charge is measured by the operator
\begin{equation}
  \hat Q = \int_{\partial\Omega} dx\, J^0_{edge}(x,t)\, .
\end{equation}
The charge of the quasi-particle created by $\psi_q$ is obtained from the commutator
\begin{equation}
  [\hat Q, \psi_{q}(\xi)] = \sqrt{\sigma}\, q\, \psi_{q}(\xi)\,.
\end{equation}
Hence, the electric charge deposited by \eqref{eq:qp} is equal to $\sqrt{\sigma}\, q$. This suggests that the vertex operators $\psi_q(\xi)$ are in one-to-one correspondence with bulk states $|(q,\lambda),z\rangle$ introduced in the previous section.

Commuting two vertex operators yields the ``statistical phase",
\begin{equation}\label{eq:statphase}
  \psi_{q_1}(\xi_1) \psi_{q_2}(\xi_2) = \psi_{q_2}(\xi_2) \psi_{q_1}(\xi_1)\, e^{\pm i\pi q_1 q_2}\,.
\end{equation}
The sign of the phase depends on the relative positions of $\xi_1$, $\xi_2$, and the starting point of the line integral (and on the homotopy classes of the paths). The conformal spin of the vertex operator \eqref{eq:qp} is given by
\begin{equation}\label{eq:confspin}
  s_q = \frac{q^2}{2}\,.
\end{equation}
These results are in accordance with the properties of the states $|(q,\lambda),z\rangle$ of the bulk TFT discussed in Sect.~\ref{chap:bulk}, for $\Delta_\lambda = 0$. In fact, the statistical phase appearing in \eqref{eq:statphase} corresponds to ``half-monodromies" in the bulk, see Eq.~\eqref{eq:monodromy}. The conformal spin \eqref{eq:confspin} coincides with the spin of the bulk state, Eq.~\eqref{eq:spin}.

We note that, under a transformation $a \mapsto a + d\alpha$, the vertex operator transforms like
\begin{equation}
  \psi_{q}(\xi)\mapsto \psi_{q}(\xi)\, e^{-2\pi i \sqrt{\sigma}\, q\, \alpha(\xi)}\,,
\end{equation}
as expected of an operator creating a particle with electric charge $\sqrt{\sigma}\, q$.

\subsection{Inter-edge tunneling and interference experiments}

\begin{figure}[!h]
\center
\includegraphics[width=\textwidth]{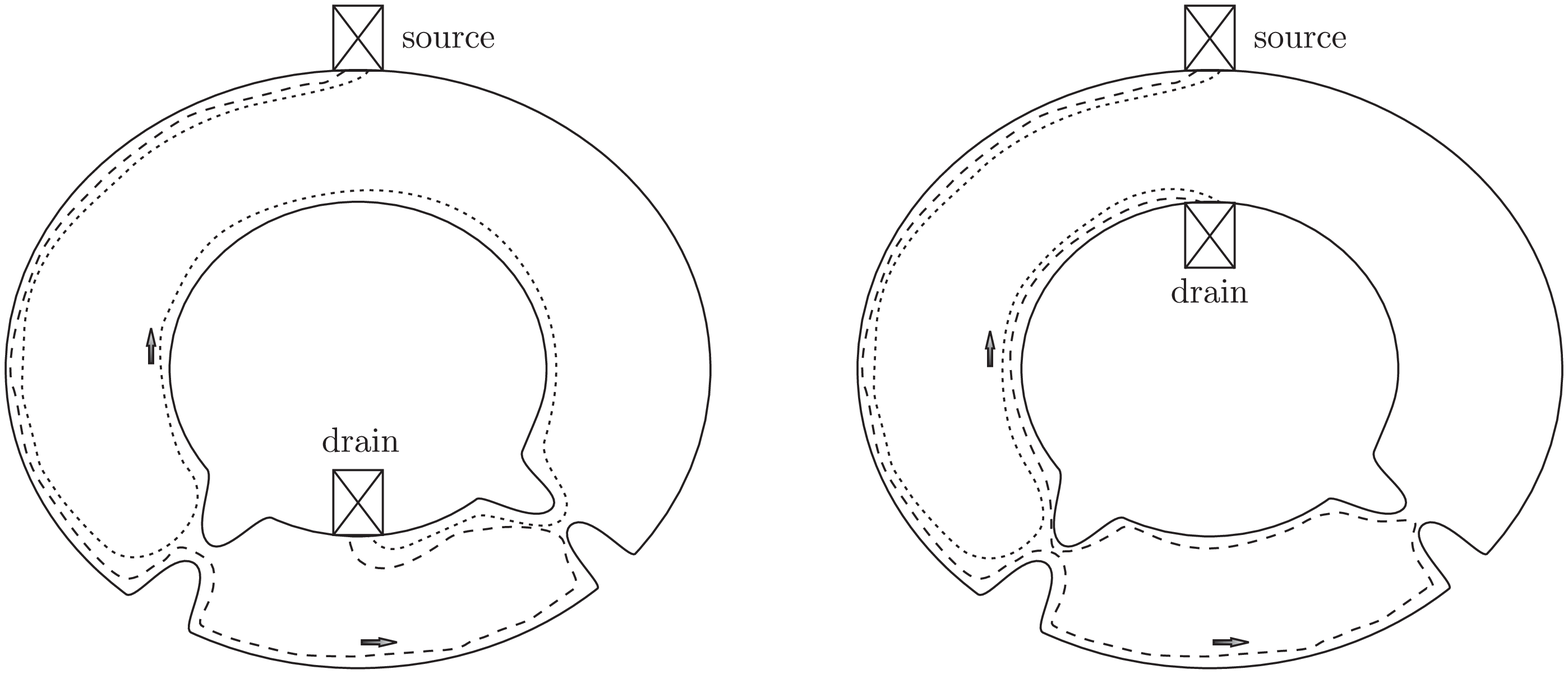}
\caption{Mach-Zehnder (left) and Fabry-P\'erot (right) interference experiments with chiral edge currents of an IHF on a Corbino-disk geometry. At two constrictions, the modes tunnel between the two edges of the sample. This plays the role of beam splitters in the optical versions of the interferometers. Properties of the quasi-particles and the magnetic flux, $\Phi$, enclosed by the loop of chiral currents lead to characteristic interference effects. Data obtained from such experiments help to constrain the set of possible effective theories describing a given IHF.
\label{fig:interference}
}
\end{figure}

The quantum field theory for the edge of an IHF can be used to predict observable effects that can be tested in beam-splitting interference experiments using electronic versions of Mach-Zehnder or Fabry-P\'erot interferometers (Fig.~\ref{fig:interference}) \cite{FrohlichPedrini2001, ChamonFreedKivelsonSondhiWen97}. For interference effects to appear, excitations need to be allowed to {\it tunnel} between different edges of the sample. This may be modeled by adding tunneling terms of the form
\begin{equation}
  V_q(x;y) \propto \int dt\, \psi_q^\dagger(x,t)\, e^{ 2\pi i\int_x^y a_1(\xi,t)\, d\xi}\, \psi_q(y,t)
\end{equation}
to the edge action.

Important phenomena are:
\begin{itemize}
  \item The tunneling current from one component of the Hall edge to another one through a quantum point contact is related to the electric charge of the particles transmitted through the contact and to the scaling dimension of the tunneling operator (exp.\ \cite{Radu08}; theory \cite{BoyarskyCheianovFrohlich09}).
  \item Aharonov-Bohm oscillations in the tunneling current are studied in~\cite{ChamonFreedKivelsonSondhiWen97,LevkivskyiBoyarskyFrohlichSukhorukov09,LFSprog}. Remarkably, they have the electronic period, $\Phi_0$, if external flux tubes are added (topological screening), but may have quasi-particle period, $\Phi_0 / q_{min}$, if edges are deformed by a modulation gate; ($q_{min}$ is the smallest fractional charge observed in the fluid).
  \item The ``visibility" of the Aharonov-Bohm oscillations as a function of bias voltage is related to the propagation speeds, $u_i$, of different channels (exp.\ \cite{BieriOberholzer09}; theory \cite{SukhorukovCheianov07,Levkivskyi08,Levkivskyi09}).
\end{itemize}

\section*{Acknowledgements}
We thank A.~Boyarsky, V.~Cheianov, I.~Levkivskyi, and E.~Sukhorukov for numerous very useful discussions. The senior author thanks C.~Glattli for having invited him to present this material at a colloquium at the Acad\'emie des Sciences in Paris. This work was supported by the Swiss National Science Foundation.


\section*{References}

\begin{thebibliography}{100}

\bibitem{FrohlichKing89}
J.~Fr\"ohlich and C.~King,
{\it The Chern-Simons Theory and Knot Polynomials},
Commun.\ Math.\ Phys.\ {\bf 126}, 167 (1989).

\bibitem{FrohlichGabbiani90}
J.~Fr\"ohlich and F.~Gabbiani,
{\it Braid statistics in local quantum theory},
Rev.\ Math.\ Phys.\ {\bf 2}, 251 (1990).

\bibitem{FrohlichKerler91}
J.~Fr\"ohlich and T.~Kerler,
{\it Universality in quantum Hall systems},
Nucl.\ Phys.\ B {\bf 354}, 369 (1991).

\bibitem{FrohlichZee91}
J.~Fr\"ohlich and A.~Zee,
{\it Large Scale Physics of the quantum Hall Fluid}, Nucl.\ Phys.\ B {\bf 364}, 517 (1991).

\bibitem{FrohlichStuder93}
J.~Fr\"ohlich and U.~Studer,
{\it Gauge invariance and current algebra in nonrelativistic many-body theory},
Rev.~Mod.~Phys {\bf 65}, 733 (1993).

\bibitem{FrohlichThiran94}
J.~Fr\"ohlich and E.~Thiran,
{\it Integral Quadratic Forms, Kac-Moody Algebras, and Fractional Quantum Hall Effect. An $ADE$-$\mathcal{O}$ Classification},
J.~Stat.~Phys.~{\bf 76}, 209 (1994).

\bibitem{FrohlichStuderThiran95}
J.~Fr\"ohlich, U.~Studer, and E.~Thiran,
{\it A Classification of Quantum Hall Fluids},
J.~Stat.~Phys.\ {\bf 86}, 821 (1995).

\bibitem{FrohlichStuderKerlerThiran95}
J.~Fr\"ohlich, T.~Kerler, U.~Studer, and E.~Thiran,
{\it Structuring the set of incompressible quantum Hall fluids},
Nucl.~Phys.~B {\bf 453}, 670 (1995).

\bibitem{FrohlichPedrini2000}
J.~Fr\"ohlich and B.~Pedrini,
{\it New applications of the chiral anomaly},
in ``Mathematical Physics 2000", A.~Fokas, A.~Grigoryan, T.~Kibble, and B.~Zegarlinski (eds.), London and Singapore: Imperial College Press 2000; arXiv:hep-th/0002195.

\bibitem{FrohlichPedrini2001}
J.~Fr\"ohlich and B.~Pedrini,
{\it Axions, quantum mechanical pumping, and primval magnetic fields},
in proc.\ of {\it Statistical Field Theory} (Como 2001), A.~Capelli and G.~Mussardo (eds.), New York and Amsterdam, Kluwer (2002);
arXiv:cond-mat/0201236.

\bibitem{FrohlichPedriniSchweigertWalcher01}
J.~Fr\"ohlich, B.~Pedrini, Ch.~Schweigert, and J.~Walcher,
{\it Universality in Quantum Hall Systems: Coset Construction of Incompressible States},
J.~Stat.~Phys.~{\bf 103}, 527 (2001).

\bibitem{BoyarskyCheianovFrohlich09}
A.~Boyarsky, V.~Cheianov, and J.~Fr\"ohlich,
{\it Effective field theories for the $\nu=5/2$ edge},
Phys.\ Rev.\ B {\bf 80}, 233302 (2009); arXiv:0904.3242.

\bibitem{LevkivskyiBoyarskyFrohlichSukhorukov09}
I.~Levkivskyi, A.~Boyarsky, J.~Fr\"ohlich, and E.~Sukhorukov,
{\it Mach-Zhender interferometry of fractional quantum Hall edge states},
Phys.~Rev.~B {\bf 80}, 045319 (2009).


\bibitem{GirvinPrange87}
R.~Prange and S.~M.~Girvin (eds.),
{\it The Quantum Hall Effect},
Springer-Verlag, New York, 1987.

\bibitem{jeanneret2004}
B.~Jeckelmann and B.~Jeanneret,
{\it The QHE as an electrical resistance standard}, Rep.~Prog.~Phys.~{\bf 64}, 1603-1655 (2001);
S\'eminaire Poincar\'e {\bf 2}, 39 (2004).

\bibitem{Klitzing80}
K.~v.~Klitzing, G.~Dorda, and M.~Pepper,
{\it New Method for High-Accuracy Determination of the Fine-Structure Constant Based on Quantized Hall Resistance},
Phys.\ Rev.\ Lett.\ {\bf 45}, 494 (1980).



\bibitem{TsuiStormerGossard82}
D.~C.\ Tsui, H.~L.\ Stormer, and A.~C.\ Gossard,
{\it Two-Dimensional Magnetotransport in the Extreme Quantum Limit},
Phys.\ Rev.\ Lett.\ {\bf 48}, 1559 (1982).

\bibitem{Laughlin81}
R.~B.\ Laughlin,
{\it Quantized Hall conductivity in two dimensions},
Phys.\ Rev.\ B {\bf 23}, 5632 (1981).

\bibitem{Laughlin83}
R.~B.\ Laughlin,
{\it Anomalous QHE: An Incompressible Quantum Fluid with Fractionally Charged Excitations},
Phys.\ Rev.\ Lett.\ {\bf 50}, 1395 (1983).

\bibitem{Halperin82}
B.~I.~Halperin,
{\it Quantized Hall conductance, current-carrying edge states, and the existence of extended states in a two-dimensional disordered potential},
Phys.\ Rev.\ B {\bf 25}, 2185 (1982).

\bibitem{Haldane83}
F.~D.~Haldane,
{\it Fractional quantization of the Hall effect: A hierarchy of incompressible quantum Hall states},
Phys.\ Rev.\ Lett.\ {\bf 51}, 605 (1983).

\bibitem{Morf86}
R.~Morf, N.~d'Ambrumenil, and B.~I.~Halperin,
{\it Microscopic wave functions for the FQH states at $\nu = 2/5$ and $2/7$},
Phys.\ Rev.\ B {\bf 34}, 3037 (1986).

\bibitem{Morf02}
R.~H.~Morf, N.~d'Ambrumenil, and S.~Das Sarma,
{\it Excitation gaps in FQH states: An exact diagonalization study},
Phys.\ Rev.\ B {\bf 66}, 075408 (2002).

\bibitem{Greiter92}
M.~Greiter, X.~G.~Wen, and F.~Wilczek,
{\it Paired Hall states},
Nucl.\ Phys.\ B {\bf 374}, 567 (1992).

\bibitem{GoldmanSu95}
V.~J.~Goldman and B.~Su,
{\it Resonant Tunneling in the Quantum Hall Regime: Measurement of Fractional Charge},
Science {\bf 267}, 1010 (1995).

\bibitem{glattli97}
L.~Saminadayar, D.~C.~Glattli, Y.~Jin, and B.~Etienne,
{\it Observation of the e/3 Fractionally Charged Laughlin Quasi-particle},
Phys.~Rev.~Lett.~{\bf 79}, 2526 (1997).

\bibitem{Umansky97}
R.~de-Picciotto, M.~Reznikov, M.~Heiblum, V.~Umansky, G.~Bunin, and D.~Mahalu,
{\it Direct observation of a fractional charge},
Nature {\bf 389}, 162 (1997);
Physica B {\bf 249 - 215}, 395 (1998).

\bibitem{JF-lectures}
J.~Fr\"ohlich, lectures in 2001 (partially based on Ref.~\cite{FrohlichGabbiani90}).

\bibitem{kitaev}
A.~Yu.~Kitaev,
{\it Fault-tolerant quantum computation by anyons}, Ann.~Phys. {\bf 303}, 2 (2003); and refs.\ therein.

\bibitem{freedman}
Ch.~Nayak, S.~Simon, A.~Stern, M.~Freedman, and S.~Das Sarma,
{\it Non-Abelian anyons and topological quantum computation}, Rev.\ Mod.\ Phys.\ {\bf 80}, 1083 (2008); and refs.\ therein.

\bibitem{currentAlgebra}
S.~Treiman, R.~Jackiw, B.~Zumino, and E.~Witten, {\it Current algebra and anomalies}, World Scientific (1985).

\bibitem{Wen2}
X.~G.~Wen,
{\it Chiral Luttinger liquid and the edge excitations in the FQH states},
Phys.\ Rev.\ B {\bf 41} 12838 (1990).

\bibitem{vafa88}
C.~Vafa,
{\it Toward classification of conformal theories},
Phys.~Lett.~B {\bf 206}, 421 (1988).

\bibitem{Difrancesco97}
P.\ Di~Francesco, P.\ Mathieu,\ and D.\ S\'en\'echal,
{\it CFT},
Springer-Verlag, New York, 1997.

\bibitem{tensorCategories}
J.~Fr\"ohlich and T.~Kerler,
{\it Quantum Groups, Quantum Categories and Quantum Field Theory},
Lecture notes in mathematics, Springer (1993).

\bibitem{FuchsRunkelSchweigert}
J.~Fuchs, I.~Runkel, and Ch.~Schweigert,
{\it Twenty five years of 2d rational CFT},
J.~Math.~Phys.~{\bf 51}, 015210 (2010), and refs.\ therein.

\bibitem{SpinOrActually07}
J.~Fr\"ohlich,
{\it Spin or, actually: Spin and Quantum Statistics},
S\'eminaire Poincar\'e {\bf 11}, 1 (2007); arXiv:0801.2724.

\bibitem{goddardOlive86}
P.~Goddard and D.~Olive,
{\it Kac-Moody and Virasoro algebras in relation to quantum physics},
Int.~J.~Mod.~Phys.~A {\bf 1}, 303 (1986).

\bibitem{MooreRead91}
G.~Moore and N.~Read,
{\it Nonabelions in the FQHE},
Nucl.\ Phys.\ B {\bf 360}, 362 (1991).

\bibitem{Wen1}
X.~G.~Wen,
{\it Non-abelian statistics in the FQH states},
Phys.\ Rev.\ Lett.\ {\bf 66}, 802 (1991).

\bibitem{ChamonFreedKivelsonSondhiWen97}
C.~de C.~Chamon, D.~E.~Freed, S.~A.~Kivelson, S.~L.~Sondhi, and X.~G.~Wen,
{\it Two point-contact interferometer for quantum Hall systems},
Phys.\ Rev.\ B {\bf 55}, 2331 (1997).

\bibitem{Radu08}
I.~P.\ Radu, J.~B.\ Miller, C.~M.\ Marcus, M.~A.\ Kastner, L.~N.\ Pfeiffer, and K.~W.\ West,
{\it Quasi-particle properties from tunneling in the $5/2$ FQH state},
Science {\bf 320}, 899 (2008).


\bibitem{LFSprog}
I.~P.~Levkivskyi, J.~Fr\"ohlich, and E.~V.~Sukhorukov, submitted to PRB; arXiv:1005.5703.

\bibitem{BieriOberholzer09}
E.~Bieri, M.~Weiss, O.~G\"oktas, M.~Hauser, C.~Sch\"onenberger, and S.~Oberholzer,
{\it Finite-bias visibility dependence in an electronic MZ interferometer},
Phys. Rev. B {\bf 79}, 245324 (2009).

\bibitem{SukhorukovCheianov07}
E.~V.~Sukhorukov and V.~Cheianov,
{\it Resonant Dephasing in the Electronic MZ Interferometer},
Phys.\ Rev.\ Lett.\ {\bf 99}, 156801 (2007).

\bibitem{Levkivskyi08}
I.~P.~Levkivskyi and E.~V.~Sukhorukov,
{\it Dephasing in the electronic MZ interferometer at filling factor two},
Phys.\ Rev.\ B {\bf 78}, 045322 (2008).

\bibitem{Levkivskyi09}
I.~P.~Levkivskyi and E.~V.~Sukhorukov,
{\it Noise-Induced Phase Transition in the Electronic MZ Interferometer},
Phys.\ Rev.\ Lett.\ {\bf 3}, 036801 (2009).

\bibitem{DeserJackiwTempleton82}
S.\ Deser, R.\ Jackiw, S.\ Templeton,
{\it Topologically massive gauge theories},
Ann.\ Phys.\ {\bf 140}, 372 (1982).

\bibitem{BelissardElstBaldes94}
J.~Bellissard, A.~van Elst, and H.~Schulz-Baldes,
{\it The noncommutative geometry of the QHE},
J.\ Math.\ Phys.\ {\bf 35} (10), 5373 (1994).

\bibitem{AvronSeilerSimon90}
J.~E.~Avron, R.~Seiler, and B.~Simon,
{\it QHE and the relative index for projections},
Phys.\ Rev.\ Lett.\ {\bf 65}, 2185 (1990); {\it Homotopy and Quantization in Condensed Matter Physics},
{\it ibid.} {\bf 51}, 51 (1983).


\end{thebibliography}
\end{document}